\documentclass[a4paper,10pt]{article}
\usepackage{mathrsfs}
\usepackage{amssymb}
\usepackage{stmaryrd}
\usepackage{txfonts}
\usepackage{authblk}
\usepackage{abstract}
\usepackage[a4paper,left=2cm,right=2cm]{geometry}
\usepackage{graphicx}

\newcommand{\be}{\begin{equation}}
\newcommand{\ee}{\end{equation}}
\newcommand{\bea}{\begin{eqnarray}}
\newcommand{\eea}{\end{eqnarray}}

\makeatletter
\newcommand\figcaption{\def\@captype{figure}\caption}
\newcommand\tabcaption{\def\@captype{table}\caption}
\makeatother

\begin{document}
\setlength{\unitlength}{1.0mm}
\title{One-loop renormalization group study of boson-fermion mixtures}
\author[1]{Boyang Liu}\affil[1]{\emph{Institute of Physics, Chinese Academy of Sciences, Beijing 100190, China}}\author[1,2]{Jiangping Hu}\affil[2]{\emph{Department of
Physics, Purdue University, West Lafayette, Indiana 47907, USA}}
\date{}
\twocolumn[\maketitle
\begin{onecolabstract}
A weakly interacting boson-fermion mixture model was investigated
using Wisonian renormalization group analysis. This model includes
one boson-boson interaction term and one boson-fermion interaction
term. The scaling dimensions of the two interaction coupling
constants were calculated as $2-D$ at tree level and the
Gell-Mann-Low equations were derived at one-loop level. We find that
in the Gell-Mann-Low equations the contributions from the fermion
loops go to zero as the length scale approaches infinity. After
ignoring the fermion loop contributions two fixed points were found
in 3 dimensional case. One is the Gaussian fixed point and the other
one is Wilson-Fisher fixed point. We find that the boson-fermion
interaction decouples at the Wilson-Fisher fixed point. We also
observe that under RG transformation the boson-fermion interaction
coupling constant runs to negative infinity with a small negative
initial value, which indicates a boson-fermion pairing instability.
Furthermore, the possibility of emergent supersymmetry in this model
was discussed.

\end{onecolabstract}
]
\section{Introduction}
Since the first observation of Bose-Einstein condensation in $^4$He
in 1995\cite{BEC}, the field of degenerate quantum gases has become
one of the most active areas of physics. Of particular interest is
the realization of boson-fermion mixtures of atom gases. They may
show very different behavior from pure fermion or pure boson gases.
Various theoretical researches have been proposed. For instance,
formation of stable strongly correlated boson-fermion
pairs\cite{Storozhenko}, instability of the mixture when there is an
attraction between bosons and fermions\cite{Miyakawa,Roth},
interspecies interactions induced attraction among
bosons\cite{Tsurumi,Pethick} and emergent supersymmetry (SUSY) from
mixtures of cold Bose and Fermi atoms\cite{Snoek,yu}. Recent
developments in atomic experiments have made it possible to realize
boson-fermion mixed gases in the laboratory. Collapse of the atomic
cloud induced by the interspecies attraction in boson-fermion
mixtures was observed experimentally\cite{Modugno}. Also, the
formation of heteronuclear Feshbach molecules has been observed in a
boson-fermion mixture of $^{87}$Rb and $^{40}$K atomic vapors in a
3D optical lattice\cite{Ospelkaus} and in an optical dipole
trap\cite{Zirbel}.

In the present work we give a renormalization group analysis on a
boson-fermion mixture model at finite temperature. Wilsonian
renormalization group approach\cite{wilson, hertz} is a popular
method to study various condensed matter problems. This technique has been
applied to a homogeneous Bose gas by several
authors\cite{Kolomeisky, Straley, Stoof}. However, it was recognized
in 1990s that the standard Wilson's momentum-shell approach must be
modified for systems involving Fermi surface\cite{Feldman, Benfatto,
Shankar1} since in such a system we renormalize not towards a single
point, the origin, but towards the Fermi surface. Renormalization
only reduces the dimension normal to the Fermi surface while the
tangential part survives\cite{Shankar2}. Besides the applications of
renormalization group in pure-boson and pure-fermion systems a RG
formalism for mixed boson-fermion systems were also discussed by
several authors \cite{Polchinski, Nayak, Onoda, Furukawa, Seiji}.

In this context one is dealing with dilute, weakly interacting
systems. This allows to effectively express the quantities of
interest in terms of a single parameter characterizing the particle
interaction. Our boson-fermion mixture model includes two important
interaction parameters $g_1$ and $g_2$ which denote short-range
boson-boson interaction and boson-fermion interaction respectively.
The renormalization group analysis shows that the scaling dimensions
of $g_1$ and $g_2$ are both $2-D$ at tree level, where $D$ is the
dimension of the system. Hence, $g_1$ and $g_2$ are both marginal
when $D=2$ and irrelevant when $D\geq3$. At one-loop level we
derived the Gell-Mann-Low equations and found that in these
equations the contributions from the fermion loops go exponentially
to zero as $\ell\rightarrow\infty$ compared with the contributions
of the boson loops. After we ignore the contributions of fermion
loops, two fixed points are found in 3 dimensional case. One is the
trivial Gaussian fixed point, the other one is the Wilson-Fisher
fixed point. At the Wilson-Fisher fixed point the parameter $g_2$
goes to zero. This implies that at one-loop level the boson-fermion
interaction of this model decouples at the critical temperature. We
also find that the the boson-fermion interaction coupling constant
with a small negative initial value runs to negative infinity under
the renormalization transformation. This could indicate a
boson-fermion pairing instability.

In the low-energy limit of a nonsupersymmetric condensed matter
system supersymmetry(SUSY) can dynamically emerge at a critical
point \cite{Lee}. For our model if the chemical potentials of boson
and fermion are equal and the two coupling constants are identical
the Hamiltonian is invariant under supergroup $U(1|1)$ \cite{Bars}.
We use RG to explore if there is such a SUSY fixed point. It turns
out in the weak interaction limit this model doesn't exhibit a SUSY
fixed point.
\section{The Model}
The model we concerned with includes one boson field $\phi$ and one
spinless fermion field $\psi$. The grand partition function can be
expressed as a functional integral, \be \mathcal {Z}=\int
D[\phi^\ast,\phi,\bar\psi,\psi]e^{-S[\phi^\ast,\phi,\bar\psi,\psi]},\ee
where \bea S[\phi^\ast,\phi,\bar\psi,\psi]=&&\int d^D x\int^\beta_0
d\tau \Big\{
\phi^\ast(\partial_\tau-\frac{\hbar^2}{2m_b}\nabla^2-\mu_b)\phi\cr
&&+\bar\psi(\partial_\tau-\frac{\hbar^2}{2m_f}\nabla^2-\mu_f)\psi
\cr &&+\frac{g_1}{2}
(\phi^\ast\phi)^2+g_2(\phi^\ast\phi\bar\psi\psi)
 \Big\}.\eea We work in D-dimensional space, where the fields depend
on spatial coordinates $\textbf x=(x_1, x_2,...x_D)$ and the
imaginary time $\tau$. In this paper we consider the cases of
$D\geq2$. The coupling constants for the short-range boson-boson
interaction and boson-fermion interaction are denoted by $g_1$ and
$g_2$.

In order to discuss the scaling of the momentum we expand the fields
in Fourier modes though \be \phi(\vec{x}, \tau)=\frac{1}{\sqrt
\beta}\sum_n\int\frac{d^Dq}{(2\pi)^D}b(\vec{q}, \omega^b_n)e^{i(\vec
q\cdot\vec x-\omega^b_n\tau)},\ee \be \psi(\vec{x},
\tau)=\frac{1}{\sqrt \beta}\sum_n\int\frac{d^DK}{(2\pi)^D}f(\vec{K},
\omega^f_n)e^{i(\vec K\cdot\vec x-\omega^f_n\tau)},\ee where
$\omega^b_n=\frac{2n\pi}{\beta}$ and
$\omega^f_n=\frac{(2n+1)\pi}{\beta}$ are the Matsubara frequencies
for boson and fermion respectively and $\beta=1/k_B T$. $k_B$
denotes Boltzmann' constant. Then we can rewrite the action in
momentum space, \bea &&S[b^\ast,b,\bar f,f]\cr
&&=\sum_n\int\frac{d^Dq}{(2\pi)^D}b^\ast(\vec{q},
\omega^b_n)(-i\omega^b_n+\epsilon_q-\mu_b)b(\vec{q}, \omega^b_n)\cr
&&+\sum_n\int\frac{d^DK}{(2\pi)^D}\bar f(\vec{K},
\omega^f_n)(-i\omega^f_n+\epsilon_K-\mu_f)f(\vec{K}, \omega^f_n)\cr
&&+\frac{g_1}{2}\cdot\frac{(2\pi)^D}{\beta}\sum_{n_1,n_2,n_3,n_4}\int\Bigg(\prod_{i=1}^4\frac{d^Dq_i}{(2\pi)^D}\Bigg)\cr
&&\Bigg\{b^\ast(\vec{q_4}, \omega^b_{n_4})b^\ast(\vec{q_3},
\omega^b_{n_3})b(\vec{q_2}, \omega^b_{n_2})b(\vec{q_1},
\omega^b_{n_1})\cr &&\delta^D(\vec q_4+\vec q_3-\vec q_2-\vec
q_1)\cdot\delta_{\omega^b_{n_4}+\omega^b_{n_3},\omega^b_{n_2}+\omega^b_{n_1}}\Bigg\}\cr
&&+
g_2\cdot\frac{(2\pi)^D}{\beta}\sum_{n_1,n_2,n_3,n_4}\int\frac{d^DK_4}{(2\pi)^D}\frac{d^DK_2}{(2\pi)^D}\frac{d^Dq_3}{(2\pi)^D}\frac{d^Dq_1}{(2\pi)^D}\cr
&&\Bigg\{\bar f(\vec{K_4}, \omega^f_{n_4})f(\vec{K_2},
\omega^f_{n_2})b^\ast(\vec{q_3}, \omega^b_{n_3})b(\vec{q_1},
\omega^b_{n_1})\cr &&\cdot\delta^D(\vec K_4+\vec q_3-\vec K_2-\vec
q_1)\cdot\delta_{\omega^f_{n_4}+\omega^b_{n_3},\omega^f_{n_2}+\omega^b_{n_1}}\Bigg\}.\eea
In above equation $\epsilon_q=\vec q^2/2m_b$ and $\epsilon_K=\vec
K^2/2m_f$ are kinetic energies for boson and fermion respectively.

\section{Renormalization Group Analysis}
\subsection{Tree Level Scaling}
We follow the Wilson's momentum-shell approach. The renormalization
group transformation involves three steps: (i) integrating out all
momenta between $\Lambda/s$ and $\Lambda$, for tree level analysis
just discarding the part of the action in this momentum-shell; (ii)
rescaling frequencies and the momenta as $(\omega, k)\rightarrow
(s^{[\omega]}\omega, sk )$ so that the cutoff in k is once again at
$\pm\Lambda$; and finally (iii) rescaling fields $\phi \rightarrow
s^{[\phi]} \phi$ to keep the free-field action $S_0$ invariant.

First Let's think about the quadratic term of the boson field. After
we integrate out a thin momentum shell of high energy mode the limit
of q (which is the radial coordinate of the momentum space) changes
from $[0, \Lambda]$ to $[0, \Lambda/s]$, where $s\gtrapprox1$. In
order to compare the action with the original one we need to rescale
the radial coordinate as \be q'=sq. \ee Hence, the cutoff in q is
back again at $\Lambda$. Here we give a definition to the scaling
dimension. If a quantity scales as \be A'=s^{[A]}A, \ee we call
$[A]$ the scaling dimension of $A$. In this manner the scaling
dimension of momentum $q$ is \be [q]=1.\ee Then the scaling
dimensions of the boson field, the energy and the chemical potential
can easily be derived from the quadratic part of the boson action.
Following the first two steps of the Wilson's renormalization group
transformation, the quadratic term of the boson action becomes \be
\sum_n\int^{\Lambda}\frac{s^{-D}d^Dq'}{(2\pi)^D}b^\ast(\vec q',
\omega'^b_n)(-is^{-[\omega^b_n]}\omega'^b_n+s^{-2}\epsilon_q')b(\vec
q', \omega'^b_n).\ee To make it invariant under the scaling
transformation we define the scaling dimension of the boson energy
as  \be [\omega_n^b]\equiv2\ee and the scaling dimension of the
boson field as \be [b]\equiv-\frac{D+2}{2}. \ee

Now we turn to the fermion case. The quadratic part of the fermion
action is given by \be S_0^f=\sum_n\int\frac{d^DK}{(2\pi)^D}\bar
f(\vec{K}, \omega^f_n)(-i\omega^f_n+\epsilon_K-\mu_f)f(\vec{K},
\omega^f_n).\ee In contrast to the boson case, low-energy modes of
fermions live near the Fermi surface. In order to preserve the Fermi
surface under scaling we can't simply scale the momentum as we did
in the bosonic case. We renormalize not towards a single point, the
orgin, but towards a surface. To make progress we define a
lower-case momentum $k\equiv|\vec K|-K_F$, which corresponds to the
low energy mode of fermions. Then it is the momentum $k$ but not
momentum $|\vec K|$ that scales under the renormalization group
transformation. Since \be
\epsilon_K-\mu_f\simeq\frac{\vec{K}^2-K^2_F}{2m}-\delta\mu_f\simeq
v_F(|\vec K|-K_F)-\delta\mu_f=v_Fk-\delta\mu_f,\ee the quadratic
part of the action can be approximated as \bea &&
\int\frac{d^DK}{(2\pi)^D}\bar f(\vec{K},
\omega^f_n)(-i\omega^f_n+\epsilon_K-\mu_f)f(\vec{K}, \omega^f_n)\cr
&&\simeq\Omega^D K^{D-1}_F\int^\Lambda_{-\Lambda}dk \bar f(k,
\omega^f_n)(-i\omega^f_n+v_Fk-\delta\mu_f)f(k, \omega^f_n),\cr
&&\eea where $v_F$ is the Fermi velocity and $\delta
\mu_f=\mu_f(T)-\mu_f(0)$ can be considered as the chemical potential
of the low-energy modes of fermions. Following the first two steps
of the renormalization group transformation this part becomes\bea
&&\Omega^D K^{D-1}_F\int^{\Lambda}_{-\Lambda}s^{-[k]}dk \bar
f(k^\prime, \omega^{\prime f}_n)(-i s^{-[\omega_n^f]}\omega^{\prime
f}_n+v_F s^{-[k]}k^\prime\cr
&&-s^{-[\delta\mu_f]}\delta\mu_f^\prime)f(k^\prime, \omega^{\prime
f}_n).\eea In order to analyze fermions and bosons in one model it
is reasonable to scale the energies of fermion and boson the same
way, that is \be [\omega_n^f]=[\omega_n^b]=2.\ee According to
Eq.(15) the scaling dimension of the low energy fermion momentum k
is the same as the fermion energy, \be [k]=[\omega_n^f]=2.\ee To
take the Eq.(15) back to the original form Eq.(14) we have to
rescale the fermion field as \be f^\prime=s^{-[k]} f.\ee Then the
scaling dimensions of the fermionic fields is \be [f]=-[k]=-2.\ee

So far we have gained the scaling dimensions of momenta, energies
and fields of both boson and fermion. Now we are ready to calculate
the scaling dimensions of the interaction coupling constant $g_1$
and $g_2$. The renormalization group transformation of the two-body
interaction terms shows more subtleties, especially for the
boson-fermion interaction term. First we study the pure boson
interaction term. After we throw away the high energy momentum
shell, the interaction becomes
\bea&&\frac{g_1}{2}\cdot\frac{(2\pi)^D}{\beta}\sum_{n_1,n_2,n_3,n_4}\int\Bigg(\prod_{i=1}^4\frac{d^Dq_i}{(2\pi)^D}\Bigg)\cr
&&b^\ast(\vec{q_4}, \omega^b_{n_4})b^\ast(\vec{q_3},
\omega^b_{n_3})b(\vec{q_2}, \omega^b_{n_2})b(\vec{q_1},
\omega^b_{n_1})\cr &&\delta^D(\vec q_4+\vec q_3-\vec q_2-\vec
q_1)\cdot\delta_{\omega^b_{n_4}+\omega^b_{n_3},\omega^b_{n_2}+\omega^b_{n_1}}\cr
&&\cdot\theta(\Lambda/s-|\vec q_4|)\theta(\Lambda/s-|\vec
q_3|)\theta(\Lambda/s-|\vec q_2|)\theta(\Lambda/s-|\vec q_1|),\cr &&
\eea where we implement $\theta$ function to generate constraints on
the momentum space instead of cutoffs in the limits of integration,
which gives a more explicit description in the scaling analysis. We
eliminate one momentum variable $\vec q_4$ using the delta function
$\delta^D(\vec q_4+\vec q_3-\vec q_2-\vec q_1)$. The above
interaction term can be written as\bea
&&\frac{g_1}{2}\cdot\frac{(2\pi)^D}{\beta}\sum_{n_1,n_2,n_3,n_4}\int\Bigg(\prod_{i=1}^3\frac{d^Dq_i}{(2\pi)^D}\Bigg)\cr
&&b^\ast(\vec q_1+\vec q_2-\vec q_3,
\omega^b_{n_4})b^\ast(\vec{q_3}, \omega^b_{n_3})b(\vec{q_2},
\omega^b_{n_2})b(\vec{q_1}, \omega^b_{n_1})\cr
&&\cdot\delta_{\omega^b_{n_4}+\omega^b_{n_3},\omega^b_{n_2}+\omega^b_{n_1}}\theta(\Lambda/s-|
\vec q_1+\vec q_2-\vec q_3|)\cr &&\cdot\theta(\Lambda/s-|\vec
q_3|)\theta(\Lambda/s-|\vec q_2|)\theta(\Lambda/s-|\vec q_1|). \eea
When the momentum $\vec q_i$ are scaled as $\vec q_i^\prime=s\vec
q_i$, the $\theta$ functions transform as \bea
\theta(\Lambda/s-|\vec q_i|)&&=\theta(\Lambda-s|\vec q_i|)\cr
&&=\theta(\Lambda-|\vec q_i^\prime|), ~~~~\mbox{for~~~~
i=1,2,3,}\eea and \bea \theta(\Lambda/s-|\vec q_1+\vec q_2-\vec
q_3|)&&=\theta(\Lambda-s|\vec q_1+\vec q_2-\vec q_3|)\cr
&&=\theta(\Lambda-|\vec q_1^\prime+\vec q_2^\prime-\vec
q_3^\prime|).\eea All the $\theta$ functions transform back to the
original forms. Then we can scale the pure boson interaction term as
\bea&&
s^{2-D}\frac{g_1}{2}\cdot\frac{(2\pi)^D}{\beta'}\sum_{n_1,n_2,n_3}\int\Bigg(\prod_{i=1}^3\frac{d^Dq'_i}{(2\pi)^D}\Bigg)\cr
&&b'^\ast(\vec q'_1+\vec q'_2-\vec q'_3,
\omega'^b_{n_4})b'^\ast(\vec q'_3, \omega'^b_{n_3})b'(\vec q'_2,
\omega'^b_{n_2})b'(\vec q'_1, \omega'^b_{n_1})\cr
&&\delta_{\omega'^b_{n_4}+\omega'^b_{n_3},\omega'^b_{n_2}+\omega'^b_{n_1}}\cdot\theta(\Lambda-|\vec
q'_1+\vec q'_2-\vec q'_3|)\cr &&\cdot\theta(\Lambda-|\vec
q'_3|)\theta(\Lambda-|\vec q'_2|)\theta(\Lambda-|\vec q'_1|). \eea
Notice that $\beta$ scales as the inverse of energy, therefore its
scaling dimension is \be [\beta]=-2.\ee In order to transform the
Eq.(24) back to its original form Eq.(20) we define \be
g'_1=s^{2-D}g_1,\ee then the scaling dimension of $g_1$ is \be
[g_1]=2-D.\ee

As discussed by R. Shankar\cite{Shankar1, Shankar2} the
renormalization group transformation of a system involving fermions
must be treated carelly. Much of the new physics stems from measure
for quartic interactions involving fermions. The boson-fermion
interaction term in our model is \bea
&&g_2\cdot\frac{(2\pi)^D}{\beta}\sum_{n_1,n_2,n_3,n_4}\int\frac{d^DK_4}{(2\pi)^D}\frac{d^DK_2}{(2\pi)^D}\frac{d^Dq_3}{(2\pi)^D}\frac{d^Dq_1}{(2\pi)^D}\cr
&&\bar f(\vec{K_4}, \omega^f_{n_4})f(\vec{K_2},
\omega^f_{n_2})b^\ast(\vec{q_3}, \omega^b_{n_3})b(\vec{q_1},
\omega^b_{n_1})\cr &&\cdot\delta^D(\vec K_4+\vec q_3-\vec K_2-\vec
q_1)\delta_{\omega^f_{n_4}+\omega^b_{n_3},\omega^f_{n_2}+\omega^b_{n_1}}\cr
&&\cdot\theta(\Lambda-| k_4|)\theta(\Lambda-|\vec
q_3|)\theta(\Lambda-| k_2|)\theta(\Lambda-|\vec q_1|).\eea First we
eliminate one variable $\vec K_4$ using the $\delta$ function
$\delta^D(\vec K_4+\vec q_3-\vec K_2-\vec q_1)$, then the
boson-fermion interaction term can be written as \bea
&&g_2\cdot\frac{(2\pi)^D}{\beta}\sum_{n_1,n_2,n_3,n_4}\int\frac{d^DK_2}{(2\pi)^D}\frac{d^Dq_3}{(2\pi)^D}\frac{d^Dq_1}{(2\pi)^D}\cr
&&\bar f(\vec{K_2}+\vec{q_1}-\vec{q_3}, \omega^b_{n_4})f(\vec{K_2},
\omega^f_{n_2})b^\ast(\vec{q_3}, \omega^b_{n_3})b(\vec{q_1},
\omega^b_{n_1})\cr
&&\cdot\delta_{\omega^f_{n_4}+\omega^b_{n_3},\omega^f_{n_2}+\omega^b_{n_1}}\theta(\Lambda-|
k_4|)\cr &&\cdot\theta(\Lambda-|\vec q_3|)\theta(\Lambda-|
k_2|)\theta(\Lambda-|\vec q_1|), \eea where \be |k_4|=|\vec K_2+\vec
q_1-\vec q_3|-K_F.\ee Functions $\theta(\Lambda-|\vec q_3|)$,
$\theta(\Lambda-| k_2|)$ and $\theta(\Lambda-|\vec q_1|)$ transform
back to their original forms in the same manner as the pure boson
case. However, the function $\theta(\Lambda-| k_4|)$ is quite
different here since $k_4$ is a function not just of $k_2$, $\vec
q_3$ and $\vec q_1$ but also of $K_F$. It's easy to check that
$\theta(\Lambda-| k_4|)$ doesn't go back to the original one after
the RG transformation. \bea
\theta(\Lambda-|k_4|)&&=\theta\Big(\Lambda-(|\vec K_2+\vec q_1-\vec
q_3|-K_F)\Big)\cr &&\rightarrow\theta\Big(\Lambda/s-(|\vec K_2+\vec
q_1-\vec q_3|-K_F)\Big)\cr &&=\theta\Big(\Lambda-s(|\vec K_2+\vec
q_1-\vec q_3|-K_F)\Big)\cr &&=\theta\Big(\Lambda-(|s\vec K_2+\vec
{q^\prime_1}-\vec {q^\prime}_3|-sK_F)\Big)\eea How can we say what
the new coupling constant is if the integration measure doesn't go
back to its old form? To solve this problem we approximate $|k_4|$
as\bea|k_4|&&=|\vec K_2+\vec q_1-\vec q_3|-K_F\cr &&=|\vec K_2+\vec
q|-K_F\cr &&\simeq K_F(|\hat{K_2}+\vec
q/K_F|-1)\cr&&=K_F(|\Delta|-1). \eea where $\vec q=\vec q_1-\vec
q_3$, $|\Delta|=|\hat{K_2}+\vec q/K_F|$ and $\hat{K_2}$ is the unit
vector of $\vec K_2$. With this approximation the transformation of
$\theta$ function is written as \bea
\theta(\Lambda-|k_4|)\rightarrow\theta(\Lambda-s|k_4|)=\theta(\Lambda-sK_F(|\Delta|-1)).\eea
Clearly for general values of $|\Delta|$ the $\theta$ function
doesn't scale invariantly. However, when \be |\Delta|=1,\ee the
$\theta$ function is invariant since
$\theta(\Lambda)=\theta(\Lambda/s)$. For the coupling constants in
condition $|\Delta|\neq1 $ we follow R.Shankar's analysis with a
soft cutoff\cite{Shankar2}: \be \theta(\Lambda-|k_i|)\approx
e^{-|k_i|/\Lambda}.\ee Then the rescaled $\theta$ function in our
boson-fermion interaction term becomes \be
\theta(\Lambda-sK_F||\Delta|-1|)\approx
e^{-sN_\Lambda||\Delta|-1|}=e^{-N_\Lambda||\Delta|-1|}e^{-(s-1)N_\Lambda||\Delta|-1|},\ee
where $N_\Lambda\equiv K_F/\Lambda$. Since $\Lambda\ll K_F$, we have
$N_\Lambda\gg1$. We can see if $|\Delta|=1$, the soft cutoff
transforms invariantly, otherwise, it doesn't matter since the
couplings will be exponentially suppressed in the limit
$N_\Lambda\rightarrow \infty$. Hence, after the scaling the
boson-fermion interaction term can be written as\bea
&&s^{2-D}g_2\cdot\frac{(2\pi)^D}{\beta'}\sum_{n_1,n_2,n_3}\int\frac{d^DK'_2}{(2\pi)^D}\frac{d^Dq'_3}{(2\pi)^D}\frac{d^Dq'_1}{(2\pi)^D}\cr
&&\bar f'(\vec{K'_2}+\vec{q'_1}-\vec{q'_1},
\omega'^b_{n_1}+\omega'^f_{n_2}-\omega'^b_{n_3})f'(\vec{K'_2},
\omega'^f_{n_2})\cr &&\cdot b'^\ast(\vec{q'_3},
\omega'^b_{n_3})b'(\vec{q'_1}, \omega'^b_{n_1})\Theta(\Lambda).\eea
Then we can identify \be g'_2=s^{2-D}g_2,\ee that is, the scaling
dimension of $g_2$ is \be [g_2]=2-D.\ee At tree level the scaling
dimensions of coupling constants $g_1$ and $g_2$ are both $2-D$.
This agrees with the reference\cite{Sachdev} for the pure boson
interaction. Hence, in 2 dimension they are all marginal.

\subsection{One-loop analysis}
In order to carry out the first step of Wilsonian renormalization
group transformation at one-loop level, we need to perform a
functional integration over the high-momentum part in the action.
For convenience we split the fields into ``slow modes" and ``fast
modes", \be \phi(\vec x,\tau)=\phi_<(\vec x,\tau)+\phi_>(\vec
x,\tau)\ee and \be \psi(\vec x,\tau)=\psi_<(\vec x,\tau)+\psi_>(\vec
x,\tau)\ee where \bea &&\phi_<(\vec{x}, \tau)=\frac{1}{\sqrt
\beta}\sum_n\int\frac{d^Dq}{(2\pi)^D}b(\vec{q}, \omega^b_n)e^{i(\vec
q\cdot\vec x-\omega^b_n\tau)}\cr
&&~~~~~~~\mbox{for}~~~~~~~~~0<|q|<\Lambda_b/s, \cr &&
\phi_>(\vec{x}, \tau)=\frac{1}{\sqrt
\beta}\sum_n\int\frac{d^Dq}{(2\pi)^D}b(\vec{q}, \omega^b_n)e^{i(\vec
q\cdot\vec x-\omega^b_n\tau)}\cr
&&~~~~~~~\mbox{for}~~~~~~~~~\Lambda_b/s<|q|<\Lambda_b,\cr &&
\psi_<(\vec{x}, \tau)=\frac{1}{\sqrt
\beta}\sum_n\int\frac{d^DK}{(2\pi)^D}f(\vec{K}, \omega^f_n)e^{i(\vec
K\cdot\vec x-\omega^f_n\tau)}\cr
&&~~~~~~~\mbox{for}~~~~~~~~~0<|K|-K_F<\Lambda_f/s^2,\cr
&&\psi_>(\vec{x}, \tau)=\frac{1}{\sqrt
\beta}\sum_n\int\frac{d^DK}{(2\pi)^D}f(\vec{K}, \omega^f_n)e^{i(\vec
K\cdot\vec x-\omega^f_n\tau)}\cr
&&~~~~~~~\mbox{for}~~~~~~~~~\Lambda_f/s^2<|K|-K_F<\Lambda_f.\eea
Then the partition function can be recast as \bea \mathcal
{Z}=&&\int
D[\phi^\ast_<,\phi_<,\bar\psi_<,\psi_<]e^{-S_<[\phi^\ast_<,\phi_<,\bar\psi_<,\psi_<]}\cr
&&\times\int D[\phi^\ast_>,\phi_>,\bar\psi_>,\psi_>]\cr &&\cdot
e^{-S_>[\phi^\ast_>,\phi_>,\bar\psi_>,\psi_>]-S_I[\phi^\ast_<,\phi_<,\bar\psi_<,\psi_<,\phi^\ast_>,\phi_>,\bar\psi_>,\psi_>]}.\eea
We next construct an effective action by integration over the fast
fields. To the one-loop order, one obtains\bea
&&e^{-S_{eff}[\phi^\ast_<,\phi_<,\bar\psi_<,\psi_<]}\cr
&&=e^{-S_<[\phi^\ast_<,\phi_<,\bar\psi_<,\psi_<]}\cr &&\cdot
\exp\Big[-\big<S_I[\phi^\ast_<,\phi_<,\bar\psi_<,\psi_<,\phi^\ast_>,\phi_>,\bar\psi_>,\psi_>]\big>_>\cr
&&+\frac{1}{2}\big<S_I[\phi^\ast_<,\phi_<,\bar\psi_<,\psi_<,\phi^\ast_>,\phi_>,\bar\psi_>,\psi_>]^2\big>_>\Big],\eea
where $\big<...\big>_>$ denotes the average over the fast
fluctuations. we perform the integrals over the fast modes by
evaluating the appropriate Feynman diagrams contributing to the
renormalization of the vertices of interest. The one-loop Feynman
graphs contributing to the renormalization are shown in Fig.1.
\begin{figure}[h]
  \includegraphics[width=8cm]{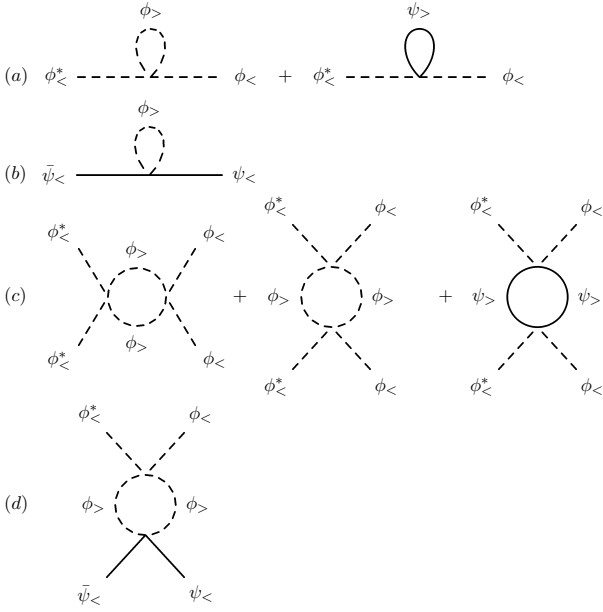}
  \figcaption{The Feynman graphs contributing to the renormalization
of (a) the boson chemical potential $\mu_b$, (b)the chemical
potential of the low-energy modes of fermions $\delta\mu_f$, (c) the
boson-boson interaction, and (d) the boson-fermion interaction.
Dashed lines denote the boson fields and solid lines denote the
fermion fields.}
 \end{figure}
After the integration over the fast fields we perform the scaling
transformations $q'\rightarrow sq$, $k'\rightarrow s^2k$, $b'(\vec
q'_i, \omega'_n)\rightarrow s^{(-D-2)/2}b(\vec q_i, \omega_n)$ and
$f'(\vec k'_i, \omega'_n)\rightarrow s^{-2}f(\vec k_i, \omega_n)$,
which bring the cutoff $\Lambda/s$ back to $\Lambda$. To keep the
action invariant under renormalization transformation one finds that
the chemical potentials and the coupling constants scale according
to the following relations up to one-loop order.\bea\mu_b\rightarrow
&& s^2\bigg\{\mu_b-2g_1\int^{\Lambda_b}_{\Lambda_b/s}
\frac{d^Dq}{(2\pi)^D}N_B(\epsilon_q-\mu_b)\cr &&-g_2\frac{\Omega^D
K^{D-1}_F}{(2\pi)^D}\int_{\Lambda_f/s^2<|k|<\Lambda_f} dk~~
N_F(v_Fk-\delta\mu_f)\Bigg\},\cr &&\eea \bea
\delta\mu_f\rightarrow~~~~~~~
s^2\bigg\{\delta\mu_f-g_2\int^{\Lambda_b}_{\Lambda_b/s}
\frac{d^Dq}{(2\pi)^D}N_B(\epsilon_q-\mu_b)\Bigg\},\eea \bea
g_1\rightarrow&&
s^{2-D}\Bigg\{g_1-g^2_1\int^{\Lambda_b}_{\Lambda_b/s}
\frac{d^Dq}{(2\pi)^D}\Big[4\beta N_B(\epsilon_q-\mu_b)\cr
&&\cdot(N_B(\epsilon_q-\mu_b)+1)+\frac{1+2N_B(\epsilon_q-\mu_b)}{2(\epsilon_q-\mu_b)}\Big]\cr
&& +g^2_2\frac{\Omega^D
K^{D-1}_F}{(2\pi)^D}\int_{\Lambda_f/s^2<|k|<\Lambda_f} dk~~\beta
N_F(v_Fk-\delta\mu_f)\cr
&&\cdot(N_F(v_Fk-\delta\mu_f)-1)\Bigg\},\eea \bea g_2\rightarrow &&
s^{2-D}\Bigg\{g_2-2g_1g_2\int^{\Lambda_b}_{\Lambda_b/s}
\frac{d^Dq}{(2\pi)^D}\beta N_B(\epsilon_q-\mu_b)\cr
&&\cdot(N_B(\epsilon_q-\mu_b)+1)\Bigg\},\eea where \bea &&
N_B(\epsilon_q-\mu_b)=\frac{1}{e^{\beta(\epsilon_q-\mu_b)}-1}\eea
and \bea
N_F(v_Fk-\delta\mu_f)=\frac{1}{e^{(v_Fk-\delta\mu_f)}+1}\eea are the
Bose-Einstein and Fermi-Dirac distribution functions which result
from the summation over the Matsubara frequencies $\omega_n^b$ and
$\omega^f_n$ and $\Omega^D$ is the D-dimensional solid angle.
setting $s=e^\ell$ and
$\Lambda_b(0)=\Lambda_f(0)=\Lambda$, we obtain the Gell-Mann-Low
equations: \bea \frac{d\mu_b}{d\ell}= &&2\mu_b-2g_1 \frac{\Omega^D
\Lambda^D}{(2\pi)^D}N_B(\epsilon_\Lambda-\mu_b)\cr
&&-g_2\frac{\Omega^D K^{D-1}_F\cdot 2\Lambda}{(2\pi)^D}
\Big(N_F(-v_F\Lambda-\delta\mu_f)\cr
&&+N_F(v_F\Lambda-\delta\mu_f)\Big),\eea \bea
\frac{d\delta\mu_f}{d\ell}= ~~~~~~~~2\delta\mu_f-g_2 \frac{\Omega^D
\Lambda^D}{(2\pi)^D}N_B(\epsilon_\Lambda-\mu_b),\eea \bea
~~~\frac{dg_1}{d\ell}=&&(2-D)g_1\cr && -g^2_1\frac{\Omega^D
\Lambda^D}{(2\pi)^D}\Big[4\beta N_B(\epsilon_\Lambda-\mu_b)\cr
&&\cdot[N_B(\epsilon_\Lambda-\mu_b)+1]+\frac{1+2N_B(\epsilon_\Lambda-\mu_b)}{2(\epsilon_\Lambda-\mu_b)}\Big]\cr
&& +g^2_2\frac{\Omega^D K^{D-1}_F\cdot 2\Lambda}{(2\pi)^D}\beta\Big[
N_F(-v_F\Lambda-\delta\mu_f)\cr &&\cdot[N_F(-v_F
\Lambda-\delta\mu_f)-1]\cr &&+N_F(v_F\Lambda-\delta\mu_f)[N_F(v_F
\Lambda-\delta\mu_f)-1]\Big],\cr&&\eea \bea \frac{dg_2}{d\ell}= &&
(2-D)g_2-2g_1g_2 \frac{\Omega^D \Lambda^D}{(2\pi)^D}\cr &&\cdot\beta
N_B(\epsilon_\Lambda-\mu_b)[N_B(\epsilon_\Lambda-\mu_b)+1],\cr
&&\eea \bea
~~~~~~~~~~~~\frac{d\beta}{d\ell}=~~~~-2\beta,~~~~~~~~~~~~~~~~~~~~~~~~~~~~~~~~~~~~~~~~~~~~~~~~~~\eea
where $\epsilon_\Lambda=\frac{\Lambda^2}{2m_b}$.  Eq.(55)
$\frac{d\beta}{d\ell}=-2\beta$ shows that for large $\ell$ the
temperature $T(\ell)$ always flows to infinity for nonzero initial
temperature. This means in the vicinity of the critical point the
Bose distribution and Fermi distribution can be reduced as \be
N_B(\epsilon_\Lambda-\mu_b)\simeq\frac{1}{\beta(\epsilon_\Lambda-\mu_b)}=\frac{k_B
T e^{2\ell}}{\epsilon_\Lambda-\mu_b},\ee \be
N_F(v_F\Lambda-\delta\mu_f)\simeq\frac{1}{\beta(v_F\Lambda-\delta\mu_f)+2}\simeq\frac{1}{2}.\ee
To absorb the factor $e^{2\ell}$ in the Eq.(56) we redefine the
scaling of the interaction coupling constants in Eq.(51)-(55) as
$g_1(\ell)=e^{(4-D)\ell}g_1$ and $g_2(\ell)=e^{(4-D)\ell}g_2$. Then
the Gell-Mann-Low equations are approximated as:\bea
\frac{d\mu_b}{d\ell}= &&2\mu_b-2g_1 \frac{\Omega^D
\Lambda^D}{(2\pi)^D}\cdot\frac{k_B T }{\epsilon_\Lambda-\mu_b}\cr
&&-g_2\cdot e^{-2\ell}\cdot\frac{\Omega^D K^{D-1}_F\cdot
2\Lambda}{(2\pi)^D} ,\eea \bea \frac{d\delta\mu_f}{d\ell}=
~~~~~~~~2\delta\mu_f-g_2 \frac{\Omega^D
\Lambda^D}{(2\pi)^D}\frac{k_B T}{\epsilon_\Lambda-\mu_b},\eea \bea
\frac{dg_1}{d\ell}=&&(4-D)g_1 -g^2_1\frac{\Omega^D
\Lambda^D}{(2\pi)^D}\cdot\frac{5k_B
T}{(\epsilon_\Lambda-\mu_b)^2}\cr && -g^2_2\cdot
e^{-4\ell}\frac{\Omega^D K^{D-1}_F\cdot \Lambda}{(2\pi)^D\cdot k_B
T},\eea \bea \frac{dg_2}{d\ell}= && (4-D)g_2-2g_1g_2 \frac{\Omega^D
\Lambda^D}{(2\pi)^D}\cr &&\cdot\frac{k_B
T}{(\epsilon_\Lambda-\mu_b)^2}.\cr &&\eea We observe that the
contributions of the fermion loops go to zero as
$\ell\rightarrow\infty$ in above equations because of the factor
$e^{-2\ell}$ and $e^{-4\ell}$. Hence, in the vicinity of the
critical point we can ignore these contributions. If we redefine the
chemical potentials and the coupling constants as \bea
&&\tilde\mu_b=\mu_b/\alpha,\cr
&&\tilde{\delta\mu}_f=\delta\mu_b/\alpha , \cr &&\tilde
g_1=g_1/\gamma,\cr &&\tilde g_2=g_2/\gamma,\eea where
$\alpha=\frac{\Lambda^2}{m}$ and $\gamma=\frac{(2\pi)^D\alpha^2}{k_B
T \Omega^D \Lambda^D}$, the the Gell-Mann-Low equations can be
further simplified as: \bea \frac{d\tilde\mu_b}{d\ell}=
&&2\tilde\mu_b-2\tilde g_1\cdot \frac{1 }{1/2-\tilde\mu_b} ,\eea
\bea \frac{d\tilde{\delta\mu}_f}{d\ell}=
~~~~~~~~2\tilde{\delta\mu}_f-\tilde g_2\cdot
\frac{1}{1/2-\tilde\mu_b},\eea \bea \frac{d\tilde
g_1}{d\ell}=&&(4-D)\tilde g_1 -\tilde
g^2_1\cdot\frac{5}{(1/2-\tilde\mu_b)^2},\cr && \eea \bea
\frac{d\tilde g_2}{d\ell}= && (4-D)\tilde g_2-2\tilde g_1\cdot
\tilde g_2\cdot\frac{1}{(1/2-\tilde\mu_b)^2}.\cr &&\eea The first
terms on the right-hand side of Eq.(63)- Eq.(66) are from the tree
level scalings. Notice that the tree level scalings of the coupling
constants $g_1$ and $g_2$ go as $4-D$ instead of $2-D$. This is
because that near a classical critical point the quantum theory
reduces to the classical theory. The same situation has been
discussed by reference\cite{Henk}.

\begin{figure}[t]
  \includegraphics[width=8cm]{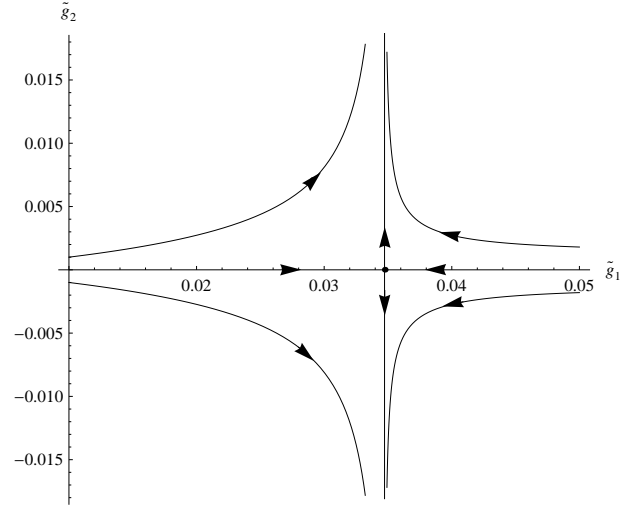}
  \figcaption{Flow diagram of the running coupling constants $\tilde g_1$ and $\tilde g_2$ in 3 dimensional case.}
 \end{figure}
For instance, we consider the 3 dimensional case. The fixed points
can be calculated as \be (\tilde\mu_b,\tilde{\delta\mu}_f,\tilde
g_1, \tilde g_2)=(0,0,0,0),\ee and\be
(\tilde\mu_b,\tilde{\delta\mu}_f,\tilde g_1, \tilde
g_2)=(\frac{1}{12},0,\frac{5}{144},0).\ee
The first one is the trivial Gaussian fixed point and the second one
is the Wilson-Fisher fixed point.
Around the Wilson-Fisher fixed point, the running of the two
coupling constants are shown in the flow diagram Fig.2. We can see
that with a small negative initial value the coupling constant
$\tilde g_2$ runs to negative infinity. This could indicate a
boson-fermion pairing instability.

\section{Conclusion}
In this paper we investigated a weakly interacting boson-fermion
mixture model by application of Wilson's renormalization group
analysis. This model includes one boson-boson interaction coupling
constant $g_1$ and one boson-fermion interaction coupling constant
$g_2$. At tree level RG analysis shows that the scaling dimensions
of $g_1$ and $g_2$ are both $2-D$. That is, the two coupling
constants are marginal in $D=2$. Here one needs to notice that the
derivation of the scaling dimension of $g_2$ is under a condition of
Eq.(34), without which we won't be able to compare the rescaled
action with the original one in RG transformation.

At one-loop level we derived the Gell-Mann-Low equations and found
that in these equations the contributions from the fermion loops
went to zero exponentially as  $\ell\rightarrow\infty$ compared with
the contributions of the boson loops. We simplify these Gell-Mann
Low equations by ignoring the fermion loop contributions and solve
for fixed points in 3 dimensional case as an example. We found two
fixed points. One is the trivial Gaussian fixed point and the other
one is the Wilson-Fisher fixed point at which $g_2$ vanishes. This
implies that the boson-fermion interaction decouples at the critical
temperature. We also drew the flow diagram of the coupling constants
$g_1$ and $g_2$ around the Wilson-Fisher fixed point. We observe
that $g_2$ goes to negative infinity with a small negative initial
value. This can be a boson-fermion pairing instability.

Supersymmetry is a symmetry that relates boson and fermion. It has
been one of the most active research areas in the high energy
physics\cite{Julius}. Various researches were also conducted to find
supersymmetry in condensed matter systems\cite{Snoek,yu}. If we have
$\mu_b=\mu_f=\mu$ and $g_1=g_2=g$ in Eq.(2), we can combine the
boson and fermion field as a doublet
$\Phi=\left(\begin{array}{c}\phi\\\psi\end{array}\right)$, which is
called superfield. Then the action can be rewrite in terms of
superfield as \bea S[\Phi^\dagger,\Phi]=&&\int d^D x\int^\beta_0
d\tau \Big\{
\Phi^\dagger(\partial_\tau-\frac{\hbar^2}{2m_b}\nabla^2-\mu)\Phi\cr
&&+\frac{g}{2}(\Phi^\dagger\Phi)^2
 \Big\}.\eea This action is invariant under supergroup
 $U(1|1)$\cite{Bars}. We used renormalization group method to
 explore if there is a supersymmetry fixed point where $\mu_b=\mu_f$ and
 $g_1=g_2$. The calculation of the Gell-Mann Low equations shows
 that our model doesn't exhibit such a fixed point.

\begin{center}\section*{Acknowledgements}\end{center} It's a pleasure to thank Professor
Wei-Feng Tsai and Dr. Chi Xiong for useful discussions.


\begin{thebibliography}{bf}
\bibitem{BEC}M. H. Anderson \emph{et al}., Science \textbf{269}, 198 (1995); K. B.
Davis \emph{et al}., Phys. Rev. Lett. \textbf{75}, 3969 (1995).
\bibitem{Storozhenko} A. Storozhenko, P. Schuck, T. Suzuki, H. Yabu and J.
Dukelsky, Phys. Rev. A \textbf{71}, 063617 (2005).
\bibitem{Miyakawa}T. Miyakawa, T. Suzuki, and H. Yabu, Phys. Rev. A \textbf{64},
033611 (2001).
\bibitem{Roth}R. Roth, Phys. Rev. A \textbf{66}, 013614 (2002).
\bibitem{Tsurumi} T. Tsurumi and M. Wadati, J. Phys. Soc. Jpn. \textbf{69}, 97 (2000).
\bibitem{Pethick} C. J. Pethick and H. Smith, \emph{Bose-Einstein Condensation in
Dilute Gases}, (Cambridge University Press, Cambridge, U.K., 2002).
\bibitem{Snoek} M. Snoek, M. Haque, S. Vandoren and H. T. C.
Stoof, Phys. Rev. Lett. \textbf{95}, 250401 (2005); M. Snoek, S.
Vandoren, and H. T. C. Stoof, Phys. Rev. A \textbf{74}, 033607
(2006).
\bibitem{yu}Y. Yu and K. Yang, Phys. Rev. Lett. \textbf{100}, 090404
(2008); T. Shi, Yue Yu, and C. P. Sun, Phys. Rev. A \textbf{81},
011604 (2010).
\bibitem{Modugno}G. Modugno, G. Roati, F. Riboli, F. Ferlaino, R. J. Brecha, and
M. Inguscio, Science 297, 2240 (2002).
\bibitem{Ospelkaus}C. Ospelkaus, S. Ospelkaus, L. Humbert, P. Ernst, K. Sengstock, and K. Bongs, Phys. Rev. Lett.
\textbf{97}, 120402 (2006).
\bibitem{Zirbel}J. J. Zirbel, K. -K. Ni, S. Ospelkaus, J. P. D'Incao, C. E. Wieman, J. Ye, and D. S. Jin, Phys.
Rev. Lett. \textbf{100}, 143201 (2008).
\bibitem{Lee} Sung-Sik Lee, Phys. Rev. B \textbf{76}, 075103 (2007).
\bibitem{Bars} I. Bars, Lect. Appl. Math. \textbf{21}, 17, (1983).
\bibitem{wilson} K. G. Wilson and J. B. Kogut, Phys. Rep. \textbf{12}, 75 (1974).
\bibitem{hertz} J. A. Hertz, Phys. Rev. B \textbf{14}, 1165 (1976).
\bibitem{Kolomeisky} E. B. Kolomeisky and J. P. Straley, Phys. Rev. B \textbf{46}, 11749
(1992).
\bibitem{Straley} E. B. Kolomeisky and J. P. Straley, Phys. Rev. B \textbf{46}, 13942
(1992).
\bibitem{Stoof} M. Bijlsma and H. T. S. Stoof, Phys. Rev. A \textbf{54}, 5085 {1996}.
\bibitem{Feldman} J. Feldman and E. Trubowitz, Helv. Phys. Acta. \textbf{63}, 156 (1990).
\bibitem{Benfatto} G. Benfatto and G. Gallavotti, Phys. Rev. B \textbf{42}, 9967 (1990).
\bibitem{Shankar1} R. Shankar, Physica A \textbf{177}, 530 (1991).
\bibitem{Shankar2} R. Shankar, Rev. Mod. Phys. \textbf{66}, 129 (1994).
\bibitem{Polchinski} J. Polchinski, Nucl. Phys. B \textbf{422}, 617 (1994).
\bibitem{Nayak} C. Nayak and F. Wilczek, Nucl. Phys. B \textbf{430}, 534 (1994).
\bibitem{Onoda} M. Onoda, I. Ichinose, and T. Matsui, Nucl. Phys. B \textbf{446}, 353
(1995).
\bibitem{Furukawa} N. Furukawa, T. M. Rice, and M. Salmhofer, Phys. Rev. Lett.
\textbf{81}, 3195 (1998).
\bibitem{Seiji} Seiji J. Yamamoto and Qimiao Si, Phys. Rev. B
\textbf{81}, 205106 (2010).
\bibitem{Henk}Henk T. C. Stoof, Dennis B. M. Dickerscheid and Koos
Gubbels, Ultracold Quantum Fields, Springer, 2009, Page 345.
\bibitem{Sachdev} Subir Sachdev, Quantum Phase Transitions,
Cambridge, 1999, page 214.
\bibitem{Julius} Julius Wess and Jonathan Bagger, Supersymmetry and
Supergravity, Princeton University Press, 1993.
\end{thebibliography}
\end{document}